\documentstyle[epsf]{mn}
\title[Mass reconstruction of A1689]
{Non-parametric mass reconstruction of A1689 from strong lensing 
data with SLAP.}
\author[Diego et al.]
  {J.M. Diego$^{1,2}$, H.B. Sandvik$^3$, P. Protopapas$^4$, M.Tegmark$^{1,2}$, N. Ben\'itez$^5$, 
   T. Broadhurst$^6$\\
   $^1$MIT Center for Space Research, Cambridge, MA 02138, USA.\\
   $^2$University of Pennsylvania. 209S, 33rd St. Department of Physics and Astronomy, Philadelphia, 
   PA 19104, USA.\\
   $^3$Max-Planck-Institut f\"{u}r Astrophysik, D-85748 Garching, Germany.\\
   $^4$Harvard-Smithsonian Center for Astrophysics, 60 Graden St. Cambridge, MA 02138, USA.\\
   $^5$Department of Physics and Astronomy, Johns Hopkins University, 3400 North Charles Street, 
   Baltimore, MD 21218, USA.\\
   $^6$School of Physics and Astronomy, Tel-Aviv University, Tel-Aviv 69978, Israel.}
\date{Draft version \today}

\pagerange{\pageref{firstpage}--\pageref{lastpage}}

\begin{document}
\maketitle

\label{firstpage}
%%%%%%%%%%%%%%%%%%%%%%%%%%%%%%%%%%%%%%%%%%%%%%%%%%%%%%%%%%%%%%%%%%%%%%%%%%%%%%%
\begin{abstract}
We present the mass distribution in the central area of the cluster
A1689 by fitting over 100 multiply lensed images with the
non-parametric Strong Lensing Analysis Package (SLAP, Diego et
al. 2004). The surface mass distribution is obtained in a robust way
finding a total mass of $0.25 \times 10^{15} h^{-1} M_{\odot}$ within
a 70'' circle radius from the central peak. Our reconstructed density
profile fits well an NFW profile  with small perturbations due to
substructure and is compatible with the more model dependent analysis
of Broadhurst et al. (2004a) based on the same data. Our estimated
mass does not rely on any {\it prior} information about the
distribution of dark matter in the cluster.  The peak of the mass
distribution falls very close to the central cD and there is
substructure near the center suggesting that the cluster is not fully
relaxed. We also examine the effect on the recovered mass when we
include the uncertainties in the redshift of the sources and in the
original shape of the sources.  Using simulations designed to mimic
the data, we identify some biases in our reconstructed mass
distribution. We find that the recovered mass is biased toward lower
masses beyond 1 arcmin (150 kpc) from the central cD and that in the
very center we may be affected by degeneracy problems. On the other
hand, we confirm that the reconstructed mass between 25'' and 70'' is
a robust, unbiased estimate of the true mass distribution and is
compatible with an NFW profile.
\end{abstract}
%%%%%%%%%%%%%%%%%%%%%%%%%%%%%%%%%%%%%%%%%%%%%%%%%%%%%%%%%%%%%%%%%%%%%%%%%%%%%%%
%%%%%%%%%%%%%%%%%%%%%%%%%%%%%%%%%%%%%%%%%%%%%%%%%%%%%%%%%%%%%%%%%%%%%%%%%%%%%%%

%%%%%%%%%%%%%%%%%%%%%%%%%%%%%%%%%%%%%%%%%%%%%%%%%%%
\section{Introduction}\label{section_introduction}
%%%%%%%%%%%%%%%%%%%%%%%%%%%%%%%%%%%%%%%%%%%%%%%%%%%
The breathtaking image of A1689 captured by the ACS camera 
onboard the Hubble space telescope (Broadhurst et al. 2004a, 
hereafter B2004) provides us with an unprecedented number 
of strong lensing arcs in a single cluster. 
The large number of arcs is due to a combination of deep multi-color imaging 
with the Hubble telescope and the inherently large Einstein radius 
of A1689. A total of 106  multiply lensed images of 30 background galaxies 
have been identified (B2004) and are spread fairly uniformly 
over an area of diameter $\sim ~300$ kpc. In principle we may obtain an estimate of the
deflection angle of the light at the location of each of the images belonging
to multiply lensed sources. This deflection relates to the projected 
gradient of the gravitational potential of the lens and hence we may derive
the surface mass density with a precision and resolution set by the 
number of multiply lensed images. 
%At the location of each lensed image 
%we may in principle obtain an estimate of the deflection of light Each individual arc contains information
%about the {\it projected} gravitational potential in a particular 
%direction, and
%The large number of arcs means we have a large 
%number of directions in which an estimate of the projected 
%gravitational field on that direction can be obtained.%need to do something here... awkward 
%the large number of arcs means we can estimate the projected
%gravitational field in large number of directions.
%%which an estimate of the projected gravitational field can be
%%obtained.
%CHEMA please check the above - i think I may have lost part of the point
%with these changes, since we have an estimate of the projected
%gravitational field onto the direction of the arc, which can be
%de-projected IF we have uniformly distributed arcs.
%
%If the arcs are uniformly distributed over the image, they 
%allow us to {\it de-project} the gravitational field and estimate 
%the mass distribution responsible for that field. 

Previous analyses of strong lensing have involved only an order of
magnitude fewer arcs per cluster and hence has not permitted the
application of a non-parametric approach, leading to only model
dependent statements in general (Kochanek \& Blandford 1991, Kneib et
al. 1993, 1995, 1996, Broadhurst et al 2000, Sand et al. 2002, Gavazzi
et al. 2004). These models have produced reliable results for simple
symmetric situations where the mass enclosed within the Einstein
radius is fairly robust to other parameters.
The quality of deep images taken with the Advanced Camera opens the way
to estimating the surface mass distribution directly without resorting to parametric
models. Non-parametric approaches have been previously explored in
several papers (Kochanek \& Blandford 1991, Saha et al. 1997, 2000,
Abdelsalam et al. 19981, 1998b, Trotter et al. 2000, Williams et
al. 2001) and more recently in Diego et al. (2004) (hereafter paper
I). In paper I, the authors showed that it is possible to
non-parametrically reconstruct a generic mass profile (with
substructure) provided the number of arcs with known redshifts is
sufficiently large.

The mass distribution of A1689 has recently been estimated using a
flexible parametric approach by B2004, who have identified over 100
background galaxies using their method. This analysis assumed a smooth
dark matter component for the bulk of the mass in the cluster plus
a small lumpy component of mass corresponding to the cluster sequence
galaxies. The cluster galaxy contribution is allowed freedom in the
ratio of M/L, but smooth component is fitted to a low order 2D
polynomial, the coefficients of which were optimized to fit the
multiply lensed systems. The model is refined as more multiply lensed
sets of images are identified by the model and incorporated to improve
its accuracy. Using this approach B2004 have been able to reliably
uncover 106 multiply lensed images of 30 background galaxies.

Non-parametric methods are interesting to explore since they provide a
model-independent estimate of the mass distribution, free of
assumptions regarding the distribution of mass in the lens
plane. Hence, this method provides a very important \emph{consistency
checks}, which should be carried out in addition to, but not
necessarily at the expense of parametric methods.
%, allowing the 
%mass to be located anywhere in the lens plane with the same probability. 
If the recovered mass distribution concurs with parametric estimates
this will add to the credibility of these results. If on the other
hand there are significant deviations, this should open the door to
interesting debates trying to understand them.  Another major
advantage of the non-parametric method is that it allows us to
estimate the systematics and errors in the recovered mass distribution
free of model assumptions. As shown in paper I, the minimization
process can take as little as a few seconds which allows for multiple
minimizations with random initial conditions. We can then study the
dispersion of the recovered solution and consequently provide an error
estimate. %though a frequentist approach

%In this paper we will use the non-parametric algorithm SLAP 
%developed by the authors (paper I). The reader can find all  
%the details in that paper where we describe the algorithm and 
%test it with simulations. Here, we will briefly summarize the main 
%ingredients to be used in this paper.
In this paper we use one of the algorithms in the Strong Lensing
Analysis Package (hereafter SLAP) developed by the authors and
introduced in paper I to reconstruct the mass distribution of A1689.
We therefore start by giving a brief summary of the main
ingredients of the method.
%%%%%%%%%%%%%%%%%%%%%%%%%%%%%%%%%%%%%%%%%%%%%%%%%%%%%%%%%%%%%%
\section{Method}\label{sec_method}
%%%%%%%%%%%%%%%%%%%%%%%%%%%%%%%%%%%%%%%%%%%%%%%%%%%%%%%%%%%%%%
\subsection{Mass-source inversion of the data}\label{sect_inversion}
The method described in this section is based on paper I, and 
the interested reader is highly encouraged to consult that paper for
the finer details. This section simply highlights the main
ingredients. 

The problem we want to solve is the inversion of the lens equation.
\begin{equation}
\vec{\beta} = \vec{\theta} - \vec{\alpha}(\vec{\theta},M)
\label{eq_lens}
\end{equation}
where $\vec{\beta}$ are the unknown positions ($\beta _x, \beta _y$) of the 
background galaxies,  $\vec{\theta}$ are the observed positions 
($\theta _x, \theta _y$) of the lensed galaxies (arcs) and 
$\vec{\alpha}(\vec{\theta})$ is the deflection angle created by the lens 
which depends on the observed positions, $\vec{\theta}$, and the unknown 
mass distribution of the cluster, $M$. The unknowns of the problem are 
then $\vec{\beta}$ and $M$.
  
Due to the (non-linear) dependency of the deflection angle, $\vec{\alpha}$, 
on the position in the sky, $\vec{\theta}$, the problem is usually regarded 
as a typical example of a non-linear problem. However, the problem
also has an equivalent formulation which can be expressed in a linear
form. The 
linearization of the problem is possible due to an observational constraint 
and a fundamental principle. 

The constraint is that the observation fixes the positions of the arcs, 
$\vec{\theta}$. 
%The problem is non-linear only in this variable, and fixing it means 
The non-linear nature of the problem is associated only with 
this variable. 
%Once $\theta$ is fixed, 
Fixing $\theta$, transforms this variable into a
constant,   
%one-to one relation with the source plane

The fundamental principle is the linear nature of the gravitational potential. 
The integrated effect of the continuous mass distribution in $\alpha$ can be 
approximated by a superposition of discretized masses. The continuous mass distribution 
can be discretized into small cells in the lens plane if the continuous mass 
distribution can be approximated by a constant over each one of the individual 
cells or in other words if the continuous mass distribution does not change much 
over the scale of the cells. This can be achieved if we divide the lens plane 
into a multiresolution grid where the size of the cell in a given position 
is inversely proportional to the mass density in that position. 

Using a multiresolution grid with $N_c$ cells, and with positions of 
the arcs, $\vec{\theta}$, fixed by observations, the problem can be 
rewritten in the linear algebraic form
\begin{equation}
\beta = \theta - \Upsilon M,
\label{eq_lens_matrix}
\end{equation}
where $\theta$ is a vector with $2N_{\theta}$ elements containing all the 
observed positions (x and y) of the $N_{\theta}$ pixels in the arcs of the 
lensed galaxy (or galaxies if there is more than one source), 
$\beta$ is made up of the corresponding $2N_{\theta}$ positions 
(x and y) of the source galaxy, $\Upsilon$ is a matrix of dimension 
$2N_{\theta}\times N_c$ where $N_c$ is the number of cells of 
the multiresolution grid used to divide the lens plane. \\

To invert the strong lensing data we use the algorithm 
of SLAP which is based on the bi-conjugate gradient method (Press et
al. 1997).  
%This algorithm was shown in paper I to be the most efficient,
%allowing for multiple inversions in a short time.
Instead of solving equation (\ref{eq_lens_matrix}) we solve the 
following  
\begin{equation}
\theta = \Gamma X.
\label{eq_lens2}
\end{equation}
Here $\Gamma$ is a matrix of dimension $2N_{\theta}\times (N_c+2N_s)$ 
and $X$ is the vector of dimension $(N_c+2N_s)$ containing all the 
unknowns in our problem, the $N_c$ cell masses, $M$, and the $2N_s$ 
central positions, $\beta _o$ (x and y), of the $N_s$ sources. 

The bi-conjugate gradient algorithm solves a system of linear 
equations,
\begin{equation}
Ax = b.
\label{eq_conj}
\end{equation}
by minimizing the function
\begin{equation}
f(x) = c - bx + \frac{1}{2}x^tAx,
\label{eq_fx}
\end{equation}
where $c$ is an arbitrary constant. When the function $f(x)$ is minimized, its gradient 
($\nabla f(x) = Ax - b$) is zero.
%\begin{equation}
%\nabla f(x) = Ax - b = 0
%\end{equation}
%That is, at the position of the minimum of the function $f(x)$
%we find the solution of equation (\ref{eq_conj}). 
The problem is formulated like this since in most cases, 
finding the minimum of equation (\ref{eq_fx}) is much easier than 
finding the solution of the system in (\ref{eq_conj}) especially when no exact 
solution exists for (\ref{eq_conj}) or $A$ does not have an inverse.

The algorithm assumes that the matrix $A$ is square. 
This does not generally hold for our case since for the matrix $\Gamma$
we typically have  $N_{\theta} >> (N_c + N_s)$. We therefore
build a new quantity called the square of the residual,$R^2$
\begin{eqnarray}
R^2 & = &(\theta - \Gamma X)^T(\theta - \Gamma X)  \\ \nonumber
    & = & 2( \frac{1}{2}\theta^T\theta -\Gamma^T\theta X + 
\frac{1}{2}X^T\Gamma^T\Gamma X).
\label{eq_R2}
\end{eqnarray}
This is clearly of the same form as equation (\ref{eq_fx}), with
$\Gamma^T\Gamma$ a square matrix. Solving the lens equation means
minimizing this quantity.
%By comparing equations (\ref{eq_R2}) and (\ref{eq_fx}) is easy to 
%identify terms, $c = 0.5\theta^T\theta$, $b = \Gamma^T\theta$ 
%and $A = \Gamma^T\Gamma$. 
The quantity $R^2$ reaches its minimum in the solution of equation 
\ref{eq_lens2} which is also solution of equation \ref{eq_conj}.
We only have to realize that;
\begin{equation}
b - AX = \Gamma^T(\theta - \Gamma X) = \Gamma^TR
\end{equation}

%The bi-conjugate gradient method finds the minimum of equation (\ref{eq_R2}) 
%by an iterative process where the number of steps (usually) is not
%higher than the dimensionality of the problem. 
%(or equivalently, the solution of equation \ref{eq_conj}) 
%by an iterative process which minimizes the function 
%$f(x)$ through a 

%The algorithm constructs two sequences of vectors $r_k$ 
%or residual and $p_k$ or search direction  %DO SOMETHING HERE
%and two constants, $\alpha_k$ and $\beta_k$. 
%by;
%\begin{equation}
%X_{k+1} = X_k + \alpha_k p_k
%\end{equation}
% and   
%chooses the residual and search direction in the first iteration to be;  
%\begin{equation}  
%r_o = p_o = b - AX_o  
%\end{equation}  
%This initial guess, $X_o$, will play an important 
%role in future sections.

The algorithm starts with an initial guess for the solution, $X_o$,
and builds an initial residual, $r_o$ and a search direction, $p_o$. 
At every iteration $k$, an improved
estimate for the residual $r_k$, the search direction $p_k$ and the 
solution, $X_k$,  is found.
The minimization is stopped when the square of the residual, $r^T
r$, is below a given value, $\epsilon$.
The beauty of this algorithm is that the successive 
minimizations are carried out in a series of orthogonal conjugate 
directions. This means it is very fast, the solution can be
found in typically 1 second of CPU time (running in a 1 GHz
processor). As we shall see below, this is crucial to allow for the
multiple minimizations required to estimate the accuracy of the
method.
%The minimization is stopped when the square of the residual, $r_k^T r_k$, 
%is sufficiently small. 

The minimization process has to be carried out through several
iterations to arrive at a division of the lens plane into a grid that
reflects well the uneven distribution of lensed images.
%This is needed since $\Gamma$ matrix depends on how
%the lens plane is divided into cells. 
For the first iteration we simply divide
the lens plane into a regular grid. After this iteration a first
estimate of the mass is used to create a new grid (and a new $\Gamma$)
where dense areas are sampled better than underdense areas. 
%In practice, each iteration is stopped when $R^2 <
%\epsilon$, with $\epsilon \rightarrow 0$.

The method has one potential pathological behavior when applied to our
problem. One can not choose the minimization threshold, $\epsilon$, to
be arbitrarily small. If one chooses a very low $\epsilon$ the
algorithm will try to find a solution which focuses the
arcs in $N_s$ sources which are $\delta$ functions. This is not
surprising as we are in fact assuming that all the $2N_{\theta}$ unknown
$\beta$s are reduced to just $2N_s$ $\beta$s, i.e the {\it point
  source solution} (see paper I).  
%The mass distribution which
%accomplishes this is usually very much biased compared to the right
%one. It shows a lot of substructure and it has large fluctuations  
%in the lens plane. 
Since the lensed galaxies are extended objects such a solution is of
course unphysical, and one therefore has to choose $\epsilon$
wisely. Since the algorithm will stop when $R^2 < \epsilon$ we 
should choose $\epsilon$ to be an estimate of the expected dispersion of 
the sources at the specified redshifts. This is the only prior which 
has to be given to the method. 
However, as shown in paper I, the specific value of $\epsilon$ is not  
critical as long as it is within a factor of a few of the true source
dispersion.  
As seen in paper I, instead of defining $\epsilon$ in terms of $R^2$, it is
better to define it in terms of the residual of the conjugate gradient
algorithm, $r_k^2$. This speeds up the minimization process
significantly.
% since we do not need to calculate the real dispersion  
%at each step but rather use the already estimated $r_k$.  
%These residuals are connected by the relation, 
%\begin{equation} 
%r_k = \Gamma^TR 
%\label{eq_rk} 
%\end{equation} 
%Imposing a prior on the sizes of the sources means that we expect the 
%residual of the lens equation, $R$, to take typical values of the order 
%of the expected dispersion  of the sources at the measured redshifts. 
%Hence we can define an $R_{prior}$ of the form; 
%\begin{equation} 
%R_{prior}^i = \sigma_i*RND 
%\end{equation} 
%where the index $i$ runs from 1 to $N_{\theta}$ and $\sigma_i$ is the 
%dispersion (prior) assumed   
%for the source associated to pixel $i$ and $RND$ is a random number 
%uniformly distributed over -1 and 1.   
%Then, we can estimate $\epsilon$ as; 
\begin{equation} 
\epsilon = r_k^Tr_k = R^T\Gamma \Gamma^TR 
\end{equation} 
As an example, 30 circular sources with 
a radius of 14 kpc located at redshifts between 1 and 6 typically 
corresponds to $\epsilon = 2.0 \times 10^{-11}$.  
%The minimization will be stopped at a point where $R^2 < \epsilon$.
%Paper I discussed how the minimization has to be stopped 
%before the absolute minimum of $R^2$ is reached. This is necessary 
%to avoid the {\it point source solution}. 

\subsection{Method Accuracy}\label{sect_accuracy}

As seen above it is crucial to stop the minimization before the
absolute minimum of $R^2$ is reached. Since we are 
minimizing an N-dimensional quadratic function ($R^2$), the area 
where we stop is an N-dimensional ellipsoid around the global minimum. 
The end point of the minimization will then vary 
depending on the initial condition, $X_o$. That is, the solution is not unique
since each minimization will stop in a different point on the $N$-ellipsoid. 
The physical meaning of this degeneracy is connected to our lack of knowledge about
the shape of the sources. When traced back to the source plane, the
pixels in the arcs are placed with any configuration within a compact
region corresponding to the size of the source. This uncertainty in
the shape of the sources can be accounted for by minimizing many
times, each time with a different initial condition, $X_o$. Using a
fast minimization algorithm like the bi-conjugate gradient is therefore
crucial in order to explore a large number of initial conditions and
estimate the scatter in the final solution.

For the current analysis the starting points for the minimization,
$X_o$, are drawn from a uniform random distribution between 0 and $1.6
\times 10^{-3} \times  10^{15} h^{-1} M_{\odot}$ for the masses and a random uniform
distribution for the $\beta$ positions in a box of 2 arcmin centered
in the cD galaxy. The value $1.6 \times 10^{-3}$  typically gives
initial total masses of around $0.5 \times 10^{15} h^{-1} M_{\odot}$
in the considered field of view. 

There are also other factors which may reduce the accuracy of the
method. One such source of uncertainty comes from the fact that the
redshifts are not known with infinite precision but have a small uncertainty. 
%Although strong lensing usually increases the total brightness 
%of a background galaxy, their high redshifts means it is
%nevertheless normal for lensed  galaxies to exhibit a low surface brightness. Measuring the 
%redshift of these distant galaxies is a challenging 
%task especially since the brightness is frequently so low that
%spectroscopic measurements are out of the question. In these
For the majority of the lensed galaxies the redshift has to be
estimated using photometric data only, and errors of 15-20\% in
redshift are quoted by B2004.

Inaccuracies in the redshifts are
problematic for our reconstruction algorithm since they propagate into
errors in the estimated angular diameter distances between us and the
source, as well as between the source and the lens. These are of course
crucial ingredients in calculating the %$\Upsilon$ 
$\Gamma$ matrix for the
linearized problem and it is therefore important that we take this
into consideration in our analysis. 

To account for the redshift uncertainty we again resort to multiple
minimizations. We use different redshift
realizations for the sources each time we solve for the lens equation
(or equivalently minimize its quadratic residual, $R^2$). This
allows us to propagate the error in the redshifts into scatter in the
solution, and gives us an estimate of the inaccuracy of the solution
through a frequentist approach.
The redshifts are generated from a Gaussian distribution, with a mean and
dispersion obtained from the data, which we assume is approximated
by a Gaussian probability distribution for simplicity.

%, we use a different redshift realization for the sources. 
%This is generated following a Gaussian distribution 
%with the mean $z_{best}$ and dispersion $\sigma$ published in B2004. 
%If the PDF for the redshift for each source was known, it could be easily used 
%to generate the random redshifts. \\

%The final ingredient adding uncertainty to our solution is the
%gridification of the lens plane. Using an adaptive grid means that
%after finding one solution, the next solution will be obtained in a
%new grid based on the old solution. 
%Each grid is based on the solution of
%the previous iteration, and is constructed so that 
%Regarding the adaptive grid, 
%the initial guess for the mass is taken
%as 0, that is, the initial grid is a regular one. 

A final source of inaccuracy in the method is the adaptive
gridding of the lens plane. 
As explained in paper I, we take the initial grid to be regular and
containing a low number of grid points. An $8
\times 8$ or $16 \times 16$ grid produces a nice initial solution
which looks roughly like a smooth version of the final solution. An
adaptive grid is then created from this first solution. It is
important that the maximum number of cells be chosen with caution. Too
few cells may not sufficiently capture the details of the mass
distributions. However, the number
of grid-cells should no be too high, exceeding the resolution set by
the projected density of the observed images.
 
%Then,
%the adaptive grid is applied over this original solution, using the
%new found solution to build a new grid. 
%%In this section we will
%%present the results obtained when we impose the adaptive grid to have
%%around 600 cells in total. 
%%In the next section we will discuss the
%%effect of taking a different number of cells in the grid but the
%%reader may intuitively predict that a very large number of cells may
%%lead to %overfeeding????
%%overfeeding problems while a very low number may not capture the
%details of the mass distribution. 

A natural upper limit for the number
of cells is 2 times the number of pixels in the data (i.e. pixels
forming part of one of the arcs) minus 2 times the number of
sources. A number of cells equal to this number would produce a square
matrix $\Gamma$. For the analysis presented in this paper, we shall see in
the next section that the number of pixels is $N_{\theta}=601$. An
estimate of the error due to choice of grid points can be obtained 
by repeating the analysis with different grid-sizes.
%shouldnt we here talk about degeneracies between  arc-pixels etc?

%A typical minimization process using 600 cells dynamical grids, changing the 
%redshifts and the initial condition, $X_o$, takes around half a minute on a 
%1 GHz processor. That is, we can find 1000 different solutions in less than 
%10 hours. A large number of minimizations allows us to explore the space of solutions 
%by including the systematics of the data (inaccurate redshifts) and the 
%algorithm (sensitivity to the initial condition and the grid). 

%%%%%%%%%%%%%%%%%%%%%%%%%%%%%%%%%%%
\section{ACS data}\label{sec_data}
%%%%%%%%%%%%%%%%%%%%%%%%%%%%%%%%%%%
\begin{figure}
   \epsfysize=8.cm 
   \begin{minipage}{\epsfysize}\epsffile{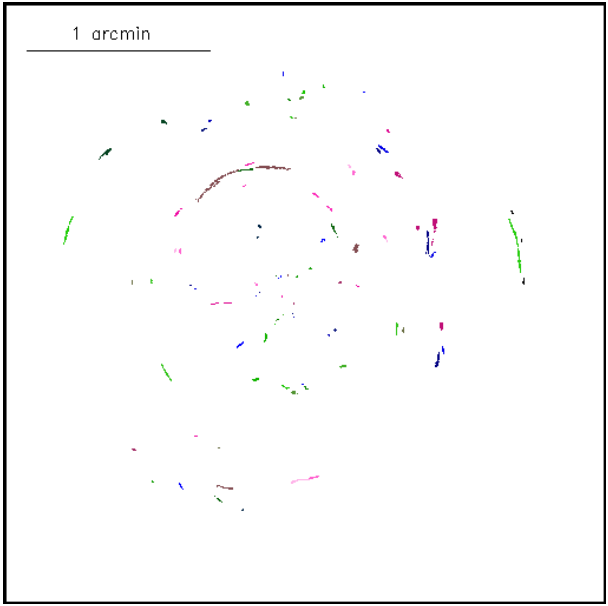}\end{minipage}
   \caption{
            Data used in the mass reconstruction. There are 106 
            arcs in this image which are assigned to 30 different sources.
            Every arc has a flag associated to the putative 
            source. Sources' redshifts range from $z \approx 1$ 
            to $z \approx 6$. The area in this plot is similar to the field 
            of view of the original data and it covers $3.3 \times 3.3$ 
            arcmin$^2$.
           }
   \label{fig_data}
\end{figure}
The data used in this paper is described in detail in B2004. Here we
only briefly summarize its main characteristics. The original ACS
image of A1689 was obtained after integration of 20 orbits  
with the Hubble space telescope in 4 bands (G,R,I and Z).  The final
published image covers a field of view of $3.3 \times 3.3$ arcmin$^2$
with a pixel size of $0.05 \times 0.05$ arcsec$^2$. The catalog with
the coordinates and redshifts of the contains the positions and
redshifts of 106 arcs, 7 of which have been spectroscopically
identified in previous works (Fort et al. 1997, Frye et al. 2002). The
bulk of the redshifts were estimated using the Bayesian
software BPZ (Ben\'itez 2000). In addition to the five bands mentioned
above, the ACS observations were complemented with U-band observations
obtained with the DuPont telescope at Las Campanas Observatory and
J,H,K data at La Silla with the NTT telescope. With these bands, the
final photometric redshifts are typically uncertain by 15-20 \%. The
106 arcs are associated with 30 systems or sources with redshifts
in the range $1<z<6$.. The positions in the catalog correspond to the center
of the arc. We only use these central positions to identify the arcs.   
Then we carefully select all the pixels in each arc to build the 
final strong lensing data set. 
We go through all the tabulated positions and select the pixels
belonging to the specified arc by eye. We only select the pixels which
are clearly connected with the arc. In the cases where the arc is too
faint, a smoothed version of the data is used to enhance the signal to
noise ratio. Eye selection is superior to algorithm selection in our case 
since software can not be trusted to separate the faintest arcs from the 
background. After all the positions in the arcs have been selected we 
repixelize the data in an area of $5 \times 5$ arcmin$^2$ using $512
\times 512$ pixels. Under this pixelization, the total number of pixels in our data set 
containing part of an arc is $N_{\theta} = 601$. The resulting data
set is show in figure \ref{fig_data}. These are the 601 $\theta$
positions which are used to invert the lens. The results are
described in the next section.

%%%%%%%%%%%%%%%%%%%%%%%%%%%%%%%%%%%%%%%%%%%%%%%%%%%%%%%%%%%%%%%%%%%%%%%%
\section{The recovered mass distribution of A1689}\label{sect_resultsI}
%%%%%%%%%%%%%%%%%%%%%%%%%%%%%%%%%%%%%%%%%%%%%%%%%%%%%%%%%%%%%%%%%%%%%%%%
In this section we present the results of our analysis when applying
the method of section \ref{sec_method} on the data from section
\ref{sec_data}. We show the results of 1000 minimizations, where
the initial mass distribution and source redshifts are randomly
varied. The maximum number of mass-cells is approximately 600. 

\begin{figure}
   \epsfysize=8.cm 
   \begin{minipage}{\epsfysize}
          \epsffile{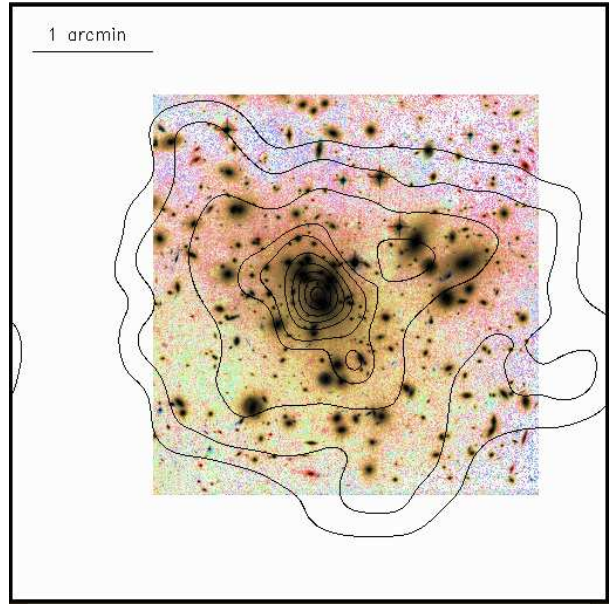}
   \end{minipage}
   \caption{
   	    Mean recovered mass (contours) compared with true ACS image. 
            The mass is the average 
            of 1000 minimizations of the lens equation where at each minimization we 
            change the grid, the initial conditions, $X_o$ and the redshifts of the sources. 
            Contours go from $0.1$ to $0.97$ times the maximum mass density  
            in intervals of $0.1$ ($0.097$ last interval).
            Total mass in the field of view is about 
            $5.2 \times 10^{14} h^{-1} M_{\odot}$.
            The field of view in this plot and the others is $5 \times 5$ arcmin$^2$ 
            unless otherwise noted. }
   \label{fig_resultMassZ1}
\end{figure}
\begin{figure}
   \epsfysize=8.cm 
   \begin{minipage}{\epsfysize}
          \epsffile{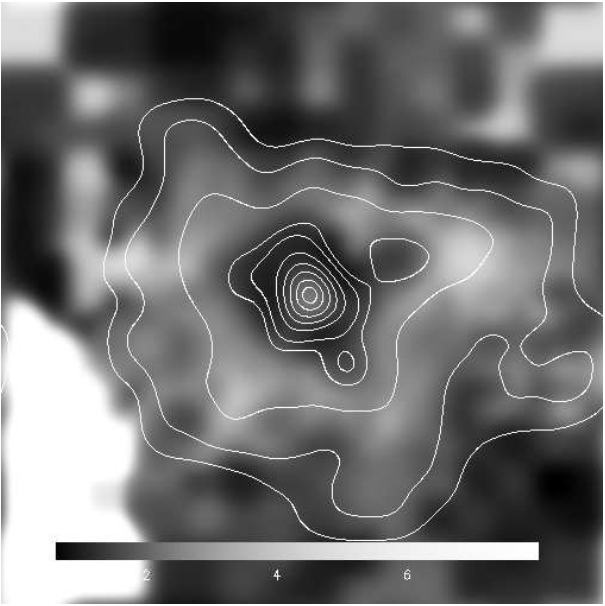}
   \end{minipage}
   \caption{
            This grey scale map shows the signal to noise ratio (SNR) of the 
            recovered mass which is obtained by dividing the mean recovered mass by 
            the dispersion of the 1000 recovered maps. For clarity, 
            the areas with SNR$>$8 have been saturated (white color). Note the low SNR at
            about 20'' from the center of mass. The contours show the mean recovered 
            mass of figure \ref{fig_resultMassZ1}
            The field of view is $5 \times 5$ arcmin$^2$.}
   \label{fig_resultSNR1}
\end{figure}

The result of this minimization process is shown in figure 
\ref{fig_resultMassZ1} where we compare the average of the 1000 
recovered solutions with the ACS optical image of A1689. 
Keeping in mind that \emph{no} information about the luminosity is
used, the first obvious conclusion from this plot is 
the existing correlation between the luminous and the 
dark matter. The peak of the mass distribution falls very close to the 
central cD galaxy. There is also a clear correlation between the position 
of the subgroup to the right and a secondary peak in the mass distribution. 
The small subgroup at $\approx 30''$ to the south of the cD seems to be sitting 
close to the top of other over-density.

The substructure within 1 arcmin of the center of the
cluster suggests that the cluster is not fully relaxed.  Another
possibility is that some of the substructure arises from
projection rather than from substructure within the main cluster.  However,
the existing correlation between the recovered mass and the galaxies
suggest that the substructure may be really present in the cluster.
Another interesting feature is that the reconstructed mass seems to be
insensitive to the external structure of A1689. There seems to be no
significant structure beyond 2 arcmins from the central cD. This can
be explained if the mass distribution beyond this radius can be
approximated by a spherical distribution.  In this case Gauss' theorem
implies that the strong lensing data should be independent of the
unknown outer mass distribution.

\begin{figure}
   \epsfysize=6.cm 
   \begin{minipage}{\epsfysize}
          \epsffile{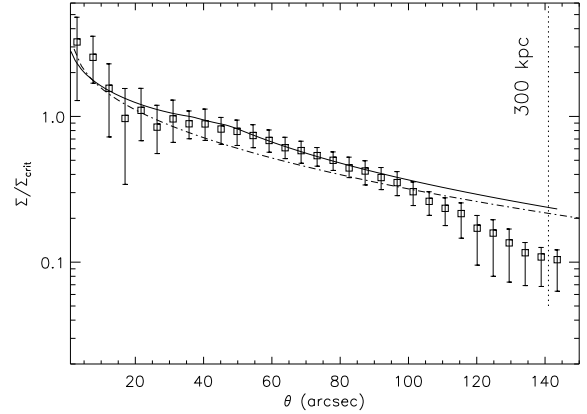}
   \end{minipage}
   \caption{
            The plot shows the mean value (squares) and the $99 \%$ confidence region of the 
            1D profiles for the 1000 minimizations in case $i)$.
            The dot-dashed line is the best fitting NFW profile found 
            in B2004. The density has been rescaled 
            by the critical density, $\Sigma_{crit}$. The thick solid 
            line is a very similar NFW profile plus an excess given by 3 
            NFW subhaloes around the main halo. See text for details. 
           }
   \label{fig_profN1}
\end{figure}

Looking at the dispersion of the 1000 minimizations tells us something about 
the reliability of our recovered mass profile.
An estimate of the dispersion of these solutions can be seen 
in figure \ref{fig_resultSNR1} where we plot the signal to noise ratio, or SNR, 
which is defined as the ratio of the mean recovered map divided by the 
standard deviation map of the solutions.
The first thing we should notice is that around 20'', the SNR drops 
below 3. In other words, the mass estimate in this region can not be trusted 
as well as in other regions. A similar behaviour can be observed at large radius
as discussed above and may imply a degeneracy set by the
limitations of the data we are using.
The insensitivity of the data to the outer regions of the mass distribution  
is suggested also when we look at the average 1D profile.
The 1D density profile is defined as the average profile 
at a given distance from the center normalized by the critical density, defined as;
\begin{equation}
\Sigma_{crit} = \frac{c^2}{4\pi G} \frac{D_s}{D_d D_{ds}} = 
4.29\times 10^{15} \frac{h M_{\odot}}{Mpc^2}
\end{equation}
In the previous expression, we have assumed that $\Sigma_{crit}$ 
is defined at the mean redshift of the sources, that is, $z=3$ (B2004).
Note that the units of $\Sigma_{crit}$ are $h M_{\odot}/Mpc^2$. 
These are the same as the recovered 
$\Sigma$ which is defined as;
\begin{equation}
\Sigma = \frac{Mass}{pixel} = \frac{h^{-1} M_{\odot}}{(h^{-1} Mpc)^2} 
= \frac{h M_{\odot}}{Mpc^2}
\end{equation}
The recovered 1D profile is shown in figure \ref{fig_profN1}. Also shown is the 
dispersion of the 1000 recovered profiles. The dotted-dashed line shows the best 
fitting NFW profile (Navarro et al. 1995) found by B2004 using the same data. 
By comparing the reconstructed profile with an NFW profile we 
can confirm the excess found in B2004. This
excess may also be well described by an NFW profile. We will discuss
this point later but here we anticipate that the data is more likely
to be compatible with an NFW profile plus an excess (thick solid line). 
However, it is clear from figure \ref{fig_profN1} that our reconstructed 
profiles differ significantly from an NFW profile at large radii thus suggesting a 
possible bias in our results here. This possibility will be explored in 
more detail later.
When we look at the normalized 1D profile (figure \ref{fig_profN1}), 
we find another striking feature which also suggests possible bias in
our results, this time in the very central region. 
As opposed to previous results based on the same data (B2004), 
the central density deviates from a NFW profile and even shows a dip at 
distances around 20 arcsec from the central peak. The same dip can be observed if 
we look at the map of the signal to noise ratio, or 
SNR (see figure \ref{fig_resultSNR1}). This may be an indication that our 
algorithm is more sensitive to tangential than radial arcs. 
The radial images contain more information about the matter distribution in the very 
center of the cluster than the tangential ones. This could be explained because we are 
minimizing the residual of the lens equation. The residual is basically dominated 
by the tangential arcs since they have more pixels than the radial arcs 
and therefore contribute more to the residual. 
Again, this possible bias will be explored later. 

Finally, an interesting conclusion from figure \ref{fig_profN1} is that using a 
non-parametric algorithm does not mean necessarily that the solution cannot be 
well constrained within the error bars. In fact, these error bars are comparable 
to the ones obtained with parametric methods.\\

%%%%%%%%%%%%%%%%%%%%%%%%%%%%%%%%%%%%%%%%%%%%%
\section{Predicted positions of the sources.}
%%%%%%%%%%%%%%%%%%%%%%%%%%%%%%%%%%%%%%%%%%%%%

\begin{figure}
   \epsfysize=8.cm 
   \begin{minipage}{\epsfysize}
          \epsffile{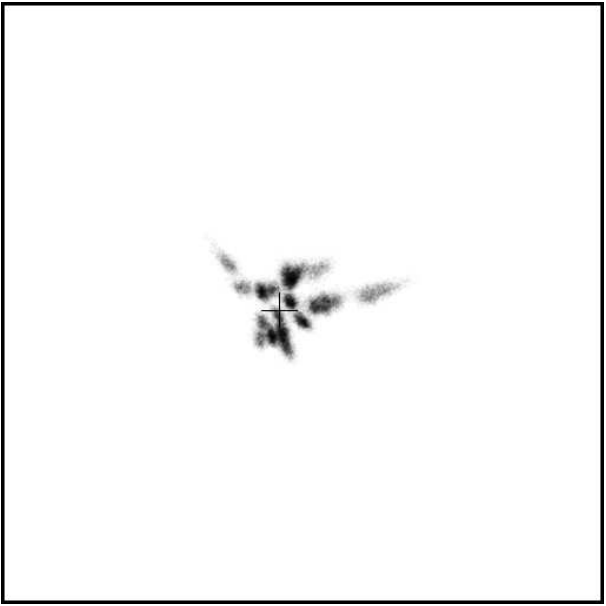}
   \end{minipage}
   \caption{
   	    Zoomed version of the recovered $\beta$ positions after 
            1000 minimizations. The field of view is $3.3\times 3.3$ arcmin. 
            The cross marks the position of the cD galaxy. Note how
            the small area of the source plane relative to the image plane 
            and implies a high magnification
            of the background galaxies with a mean value of $\sim 8$.
% XXX  this is possible because I have checked each set of images in
%detail to see that their colours and internal lumps and bumps all match up - you can
%see how uniquely well we fit each image in the lower panel of the sets of postage stamps
% that I calculated in B2004 - this model check on the appearence of each image
% is essential for checking the validiy of the identifications and more importantly if
%other images appeared at other locations and do not correspond to an observed object
%then of course there would be something very fishy going on and I would see that.
%so I am really sure that each set of images is unique and correct! If you want me
%to take your 2D deflection field I can check out your quesion  by inputting a 
% given image , and relensing to se if the other images appear close to plausible
% counterimages that I have overlooked?
%            This plot suggest that 
%            the number of sources may be smaller than the assumed 30.
           }
   \label{fig_RecovB}
\end{figure}

The solution found in the previous section also gives us the original
position of the sources. Let us recall that in our algorithm we assume
that the sources are point-like and they are described by just two
numbers, namely the $x$ and $y$ coordinates at the center.  For each
of the 1000 minimizations we obtain an estimate of the $(x,y)$
position of each source. The result is plotted in figure
\ref{fig_RecovB}.  The recovered sources fall in a small area of
$\approx 1 \times 1$ arcmin$^2$. Some sources seem to fall on top of
others. Given the uncertainties in the photometric redshifts, it could
happen that some of the sources are at the same redshift. This,
together with the fact that they appear in the same area in the sky,
make us think that some of these sources may be the same.  We should
note however that previous work has identified a systematic problem
when minimizing the lens equation in the source plane, namely the fact
that the minimization is biased toward higher masses for the lens and
with the sources being in a more compact region. If we are indeed
affected by this, this would explain why the sources seem to fall in
such a compact region. This possible systematic effect will be also
studied later.
%%%%%%%%%%%%%%%%%%%%%%%%% 
\section{Critical curves} 
%%%%%%%%%%%%%%%%%%%%%%%%%%%% 
\begin{figure}
   \epsfysize=8.cm 
   \begin{minipage}{\epsfysize}
          \epsffile{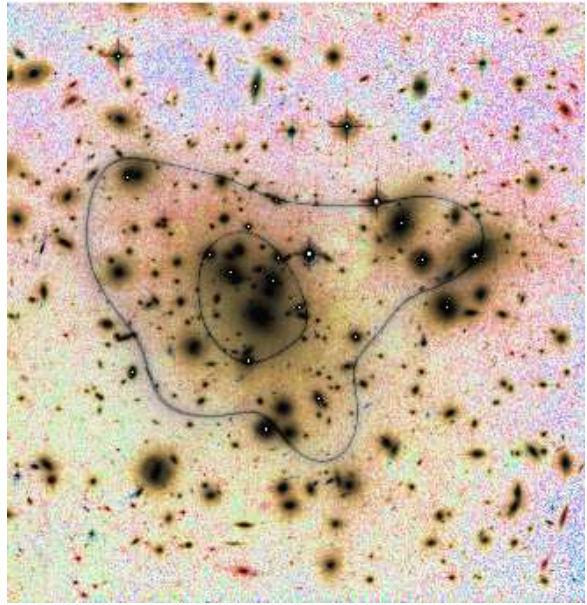}
   \end{minipage}
   \caption{
            Critical curves for the mean recovered mass in 
            figure \ref{fig_resultMassZ1}. The field of view is the same as the 
            original image (3.3 arcmin). Note the clear formation of a radial
            critical curve whose size relative to the tangential critical curve
            requires a shallow central mass profile.
           }
   \label{fig_RecovM2D_V2}
\end{figure}
It is interesting to look at the critical curves of our reconstructed 
mass. These curves are defined as the regions where the magnification
diverges. Normally one expects to see two kind of curves, the tangential 
critical curve and the radial critical curve. The first one is normally 
associated with the Einstein radius and is where the big radial arcs tend to 
appear. 
%The distance at which the tangential critical curve appears 
%typically scales as the square root of the mass embedded within the big 
%tangential arcs.  IS THIS ARGUMENT NOT CIRCULAR?
%The distance at which the tangetial critical curve appears typically
%scales as the square root of the mass 

The radial critical curve defines the region where two multiple images
merge or split in the radial direction. This curve is very interesting
because it is sensitive to the particular profile of the inner region
of the cluster.  If we change the total mass, the concentration
parameter and the characteristic scale, $r_s$, such that the
tangential critical curve does not change much (i.e, we do not change
the mass embedded within the giant tangential arcs) then we observe
that smaller $r_s$'s produce smaller radial critical curves.  In other
words, the ratio between the tangential and the radial critical curves
tells us something about the steepness of the profile between the
radii of the giant arcs and the center. A steep profile will produce a
small relatively small radial critical curve, for a fixed tangential
critical curve.  A relatively large radial critical curve, is
generated by a flatter profile near the center of the cluster. Note that for
profiles steeper than the isothermal case, the radial
critical curve is reduced to a point at the position of the lens.

Previous analysis of A1689 based on the same data (B2004) found a
relatively large radial critical curve extending up to 20'' from the
center of the cluster. NFW profiles are compatible with these large
radial critical curves only if the halo characteristic radius, $r_s$,
is relatively large. B2004 found best fitting values of $r_s = 310
h^{-1}$ kpc and concentration parameter $C_N = 8.2$ (with $C_N =
R_{virial}/r_s$). An NFW profile like this one reproduces well the
derived critical curves in B2004.

%  In a second paper, Broadhurst et
%al. (2004b), the authors combined the strong lensing ACS data with
%weak lensing of deep, wide field SUBARU images and found that the best
%fitting NFW model for the combined data has a surprisingly high
%concentration parameter ($C_N = 13.7$).  Assuming a virial radius of $
%R_{virial} < 2 h^{-1}$ Mpc would imply that the characteristic radius
%should be $r_s < 140 h^{-1}$ kpc.  Such a small characteristic radius
%would have an impact on the radial critical curve. It should be
%smaller than the one corresponding to $r_s = 300 h^{-1}$ kpc (assuming
%the total mass within the giant arcs is fixed).

%XXX we have redone this a bit and the r_s has now comporsated a bit for the higher
%C and the radial critical curve is not quite as bad comes out at about 15'' instead of
%~20 observed - so I'd say that was sucha bad agreement

The critical curves of our mean recovered model (see figure
\ref{fig_resultMassZ1}) are shown in figure \ref{fig_RecovM2D_V2}.  By
comparing with the critical curves in B2004 we see that the inner
curve (radial critical curve) is similar (or even larger in some
areas) than the one obtained in B2004. This fact suggests that the
characteristic scale, $r_s$, must be indeed large, of the order of 300
$h^{-1}$ kpc or more.  Also from the same plot, our critical curves
show a smoother behavior than previous analysis (B2004), which may
suggest that we are not very sensitive to small details in the mass
distribution.  More specifically, the differences between our
recovered critical curves and the ones found in B2004 are bigger in
the case of the radial critical curve which is more sensitive to the
details in the central part of the cluster. A higher resolution is
expected in the center for the modeling of B2004 because of the mases
of the tight clump of luminous cluster galaxies found there are
included in the model as part of the cluster sequence component
(B2004). This level of detail which is not easy to reproduce in detail
with our fully non-parametric model, which would require more
constraints in the center for a more detailed fit here, hence our results in the
center $r<20''$ should probably be regarded as a somewhat smoothed
version of the central mass profile.
This very last point may be connected with the drop interior to the critical 
curve (around 20'' from the center) in the mass density profile 
(see figure \ref{fig_profN1}).  This feature in the profile could be
due to a degeneracy among 
the masses in the cells in the very central region of the cluster and could be
easily  %%WHAT DO WE MEAN HERE? DEGENERACY BETWEEN WHAT MODELS?
explained by the argument used above that our algorithm is less sensitive 
to the radial than to the tangential arcs. 
The features in the profile may be real or due to a systematic bias in our algorithm. 
Answering this question is the purpose of the next sections.

%%%%%%%%%%%%%%%%%%%%%%%%%%%%%%%%%%%%%%%%%%%%%%%%%%%%%%%%%%%%%%%%%%%%%%
\section{Error analysis and possible systematics}\label{sect_systemI}
%%%%%%%%%%%%%%%%%%%%%%%%%%%%%%%%%%%%%%%%%%%%%%%%%%%%%%%%%%%%%%%%%%%%%%
The results in the previous section offer some answers about the 
mass distribution in A1689 but also raise some serious questions 
about the reliability of our results. A visual comparison with 
the results of B2004 indicates some disagreement 
between our mass distribution and theirs. Our recovered mass 
distribution shows substructure within the 
central 200 $h^-1$ kpc ($1.5$ arcmin) with dips and peaks around the 
central peak. The overall mass distribution is similar in shape to that of
B2004, but with more pronounced substructure.
The difference can be partially explained by the fact that parametric
methods implicitly assume a smooth distribution for the main dark
matter component with no dips while we do not.  The second possibility
is that the dips are an artifact coming from degeneracies of the
modelling procedure.  As shown in paper I, one may expect a variety of models to be
consistent with the data.  Some of these models may show degeneracies
between neighboring cells at small scales if the result is not
sensitive to these small scales, although in general is not possible
to predict where the degeneracies will appear.  One expects that the
range of valid models reduces as the number of arcs increases. This
means that each case has to be studied separately. This possibility
will be explored further in the next section.\\

In this section we focus on another source of systematics.  In section
\ref{sect_resultsI} we included in our analysis the numerical
uncertainties in our algorithm. They were the uncertainty in the
knowledge of the redshift of the sources and the uncertainty in the
shape of the sources.  The uncertainty in the redshift was included by
assigning different redshifts to the sources at each minimization
(Gaussian distribution), while the uncertainty in the shape of the
sources was included by minimizing many times, each one with a
different initial condition, $X_o$. \\

In section \ref{sect_resultsI} we also changed the grid at each
iteration using our dynamical adaptive grid which constructs the new
grid based on the previous solution. For doing that we had to fix one
parameter of the algorithm, the total number of cells, $N_c$. The
algorithm needs another parameter to be defined, namely the minimum
residual we want to achieve, $\epsilon$.  The algorithm stops when
$R^2 < \epsilon$, where $\epsilon$ can be defined by the size of the
sources and their number. 
In sections \ref{sec_method} and \ref{sect_resultsI} we gave
some intuitive motivation on how to choose $\epsilon$ and $N_c$
respectively.  In this section we address the issue of how sensitive
the results are to these two parameters.

We consider three different scenarios or cases.\\ $\bullet$ Case $i)$,
the minimization is performed with a number of cells $N_c \approx 600$
and $\epsilon = 2 \times 10^{-11}$.  This is the case used to present
the results in section \ref{sect_resultsI}.\\ $\bullet$ Case $ii)$, as
in case $i)$ but we reduce the number of cells to $N_c \approx 300$.\\
$\bullet$ Case $iii)$, as in case $i)$ but we reduce the size of the
sources to $\epsilon = 5 \times 10^{-12}$\\
\begin{figure}
   \epsfysize=6.cm 
   \begin{minipage}{\epsfysize}
          \epsffile{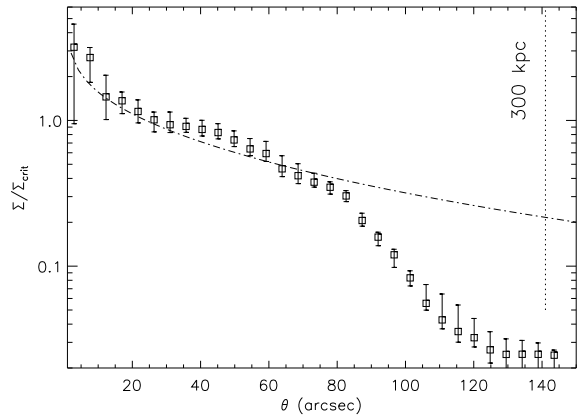}
   \end{minipage}
   \caption{
            Recovered 1D profile with error bars (at 99 \% level). 
            The dot-dashed line is the best NFW profile found in 
            B2004. This is case for ii) (300 cells, $2E-11$).
           }
   \label{fig_profM3}
\end{figure}
Case $i)$ was already studied in the previous sections and is used
here for comparison.  For each of the cases $ii)$ and $iii)$ , we run
another 1000 minimizations changing the starting point, $X_o$, the
redshifts and the grid as we did in case $i)$ (section
\ref{sect_resultsI}).

\begin{figure}
   \epsfysize=8.cm 
   \begin{minipage}{\epsfysize}
          \epsffile{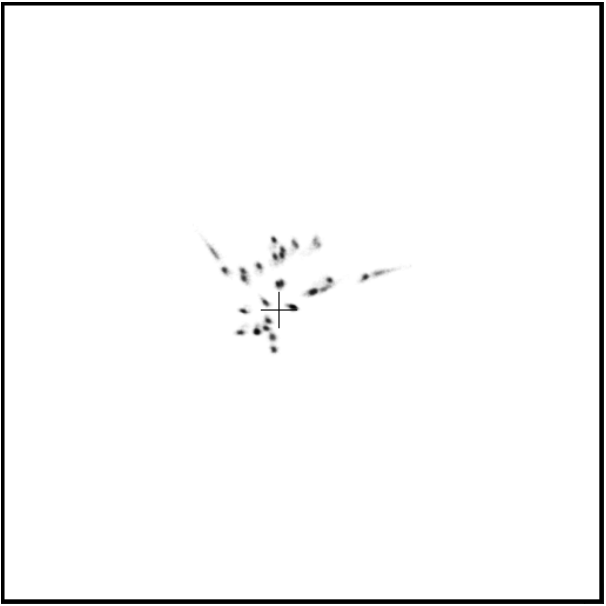}
   \end{minipage}
   \caption{
   	    Zoomed version of the recovered $\beta$ positions after 
            1000 minimizations for case $ii)$. Field of view is $3.3 \times 3.3$ arcmin$^2$.
           }
   \label{fig_RecovB3}
\end{figure}

In case $ii)$, by reducing the number of cells we reduce the number of
possible solutions, i.e we reduce the uncertainty in the solution. We
also degrade the resolution since we have to fill the same space ($5
\times 5$ arcmin$^2$) with half the number of cells. After averaging
1000 minimizations, the recovered mass distribution\footnote{Figure
available in http://darwin.physics.upenn.edu/SLAP/} looks similar to
the one found in case $i)$ with the main difference being in the outer
areas were case $ii)$ shows an even larger deficit in mass when
compared to the NFW profile.  The critical curves$^{\star}$ also look
very similar to the ones found in case $i)$ but showing a slightly
larger radial critical curve which suggests a higher concentration of
mass near the center of the cluster. The average of the 1D profiles
together with its $99 \%$ error bars can be seen in figure
\ref{fig_profM3}. The plot clearly demonstrates the departure from the
NFW profile at large radii.  It also shows the reduction in the
dispersion of the solutions as well as a lack of a dip at 20''.  The
same effect can be seen when we look at the predicted position of the
sources (figure \ref{fig_RecovB3}). Contrary to what happened in case
$i)$ (see figure \ref{fig_RecovB}), the predicted positions of the
sources in case $ii)$ do not suggest a smaller number of sources. A
closer look reveals that in case $ii)$ the smaller number of cells
produces a sequence of grids with very small differences between
them. In other words, in case $ii)$ we are in a situation in which the
grid has been practically fixed from iteration 1. This fact
contributes crucially to the reduction in the range of solutions
(masses and $\beta$ positions). \\
\begin{figure}
   \epsfysize=8.cm 
   \begin{minipage}{\epsfysize}
          \epsffile{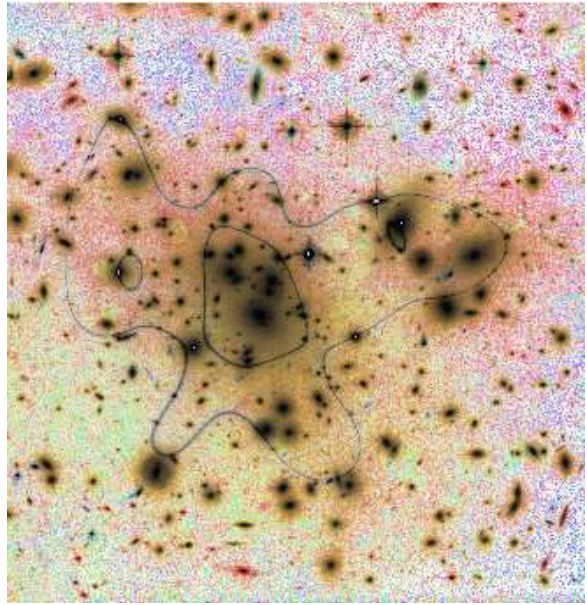}
   \end{minipage}
   \caption{
            Critical curves for the mean recovered mass in 
            the case $iii)$. The field of view is (3.3 arcmin)$^2$ 
           }
   \label{fig_CritCurv2}
\end{figure}
Case $iii)$ is interesting to explore because it forces the algorithm
to find a solution closer to the unphysical {\it point source
solution}. The total dispersion in the source plane has now been
reduced by a factor 4. The solutions achieve this by adding more
substructure to the mass distribution and when $\epsilon$ is made
small enough, the $\beta$ positions are also shifted toward the
position of the center of mass. This effect is well known and it was
studied in paper I.  In our particular case, the mean mass
distribution of the 1000 solutions looks again similar$^{\star}$ to
the one found in case $i)$ but showing more substructure.  The average
1D profile$^{\star}$ is also similar to the one in figure
\ref{fig_RecovB}.  Here we present only the critical curves in figure
\ref{fig_CritCurv2} were the effect of the extra substructure can be
appreciated. \\ The residual, $R$ or $\epsilon$, has the physical
meaning of being the variance or size of the sources.  Setting a very
small $\epsilon$ produces a biased mass distribution which focuses the
arcs into very small sources or point sources. The {\it point source
solution} achieving this is normally unphysical as it was shown in
paper I.  On the contrary, choosing a large $\epsilon$ will stop the
minimization early, resulting in a {\it short sighted cluster},
meaning the solution cannot focus the arcs properly. This short
sighted cluster solution is normally a smoother, lower-mass version of
the real solution.

%%%%%%%%%%%%%%%%%%%%%%%%%%%%%%%%%%%%%%%%%%%%%%%%%%%%%%%%%%%%%%%%%%%%%%
%\section{Validating the conclusions with simulations}\label{sect_sims}
\section{Testing the results with simulations}\label{sect_sims}
%%%%%%%%%%%%%%%%%%%%%%%%%%%%%%%%%%%%%%%%%%%%%%%%%%%%%%%%%%%%%%%%%%%%%%

The previous section has two possible interpretations.  On the
pessimistic side, we raised concerns about the reliability of our
results since we show how the results change depending on our choice
of number of cells and the stopping point of the minimization.  On the
other hand, the positive interpretation is that the change in the
results is not dramatic and our conclusions seem to be relatively
insensitive to big changes in the minimization process.
%There are however pending questions regarding  
%how confident we are in our recovered mass structure. Ideally we would like 
%to know what we can trust and what we can not. 

Although the last section gave us an idea about the dispersion in the
solution, it did not address the issue of whether or not the recovered
solution is \emph{biased}.  The problem in answering this question is
of course that we do not know what the real mass distribution is, thus
there is nothing to compare our results with.  The aim of this section
is to rectify this by using a simulated data set which mimics the main
features of the real data.  With a simulation we can easily check
aspects like how sensitive the data is to the mass distribution in the
very center or in the area beyond the tangential arcs.  Our simulated
cluster is a simplified version of the recovered mass distribution,
made up of a superposition of NFW profiles.
%By {\it mimicking} the data we mean a simulation which has more or less the same 
%characteristics of the real data. DON'T NEED A DEFINITION OF MIMICKING;)
%%There are several ways of doing this. In our case we will follow a simple process. 
%%The mass distribution will be generated from a superposition of NFW
%%profiles. 
%In order to account for the possible missing mass in the outer parts of the 
%recovered cluster, the total mass of our simulated cluster will be larger than the 
%total mass found in section \ref{sect_resultsI}, but with a similar mass within 
%the giant arcs. 
Since the recovered solution has a mass deficit in the outer parts, the
simulated cluster has a larger total mass, but is chosen so that it
resembles well the mass distribution within the giant arcs.

\begin{figure}
   \epsfysize=8.cm 
   \begin{minipage}{\epsfysize}
          \epsffile{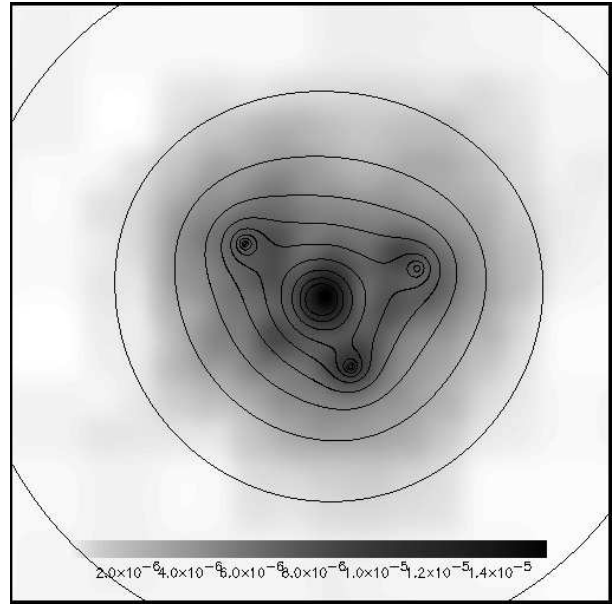}
   \end{minipage}
   \caption{
            Recovered mass (grey scale) compared with the original simulated one
            (contours). The contours increase in steps of 0.5 times the maximum central 
            density starting at 0.05 times the maximum.
            The units of the greyscale map are $10^{15} h^{-1} M_{\odot}/pixel$ and there are 
            $512^2$ pixels in the image.
            The field of view is (5 arcmin)$^2$  }
   \label{fig_MassSim1}
\end{figure}

We use a total mass of $M_T = 0.68 \times 10^{15} h^{-1} M_{\odot}$ in
the field of view ($5 \times 5$ arcmin$^2$).  For simplicity, our
simulated cluster is made of only four NFW halos. The main halo is
assigned a mass of $M_1 = 0.53 \times 10^{15} h^{-1} M_{\odot}$ and
placed at the maximum of the averaged recovered mass in section
\ref{sect_resultsI}.  The second halo is given a mass of $M_2 = 0.07
\times 10^{15} h^{-1} M_{\odot}$ and centred in the northeastern
subgroup. The third halo with $M_3 = 0.03 \times 10^{15} h^{-1}
M_{\odot}$ is centered to the south-east of the main group, and
finally the fourth halo with $M_2 = 0.05 \times 10^{15} h^{-1}
M_{\odot}$ is placed to the north-west of the main halo (see figure
\ref{fig_MassSim1}).  This simulated cluster resembles the
reconstructed mass profile found in section \ref{sect_resultsI} but
with the difference that it has a sharp cusp in the center (plus 3
off-peak sharp cusps) and the tails of the distribution do not fall
off as quickly as in the recovered mass distribution. We have also
verified that the model roughly reproduces the recovered critical
curves$^{\star}$. The 1D profile of this simulated cluster is shown in
figure \ref{fig_profN1} (thick solid line) where it is compared with
the reconstructed 1D profile and the best fitting NFW profile of
B2004.  The projected mass distribution is shown in figure
\ref{fig_MassSim1}.  For the lensing simulation the cluster is located
at the same redshift as A1689 ($z = 0.18$).

The second ingredient of the simulation are the sources.
%For this, we use 30 sources and we 
%place them at the published redshifts in B2004. 
We use 30 sources whose $\beta$ positions are taken as random within a
box of $1 \times 1$ arcmin around the center of the main halo.  The
sources are assumed to be circular with radii of a few kpc, and are
placed at the redshifts published in B2004.

The final part of the simulation is to find the arcs corresponding to
the previous configuration$^{\star}$. For this we use a simple ray-tracing algorithm. 
With this simulated data we follow a reconstruction process similar to
the case $i)$ in the previous section. 
We run 1000 minimizations (but with 500 cells instead of 600 and with 
$\epsilon = 2.0 \times 10^{-11}$) 
and calculate the mean value and dispersion of the solutions.
\begin{figure}
   \epsfysize=6.cm 
   \begin{minipage}{\epsfysize}
          \epsffile{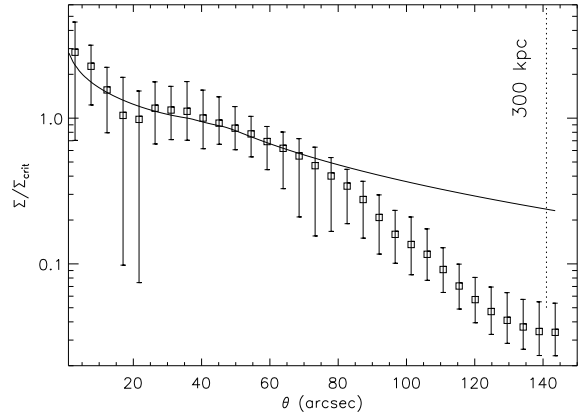}
   \end{minipage}
   \caption{
            The mean and error bars of the 1000 recovered profiles after 
            changing the initial conditions, $X_o$, the redshifts and the grid at 
            each minimization. The thick solid line is the original profile of the 
            simulated cluster.
           }
   \label{fig_MassSim1_prof}
\end{figure}

The average of the 1000 recovered masses is shown in figure
\ref{fig_MassSim1} as a grey scale map and it is compared with the
original mass distribution (contours).  The position of the main halo
is reconstructed with good accuracy. In the position of the secondary
halos we reproduce an over-density although a spurious over-density
also appears in the south-west of the main halo. The total recovered
mass is $4 \times 10^{15} h^{-1} M_{\odot}$, that is 40 \% smaller
than the original total mass. This deficit in mass is again
concentrated in the outer areas, beyond the position of the giant arcs
as can be seen from the recovered 1D profile (figure
\ref{fig_MassSim1_prof}). The simulation confirms that the algorithm
is insensitive to the mass distribution beyond the most distant arcs
from the centre.  Another interesting conclusion from figure
\ref{fig_MassSim1_prof} is that the algorithm also seems to have some
problems finding the right mass in the central region. It
over-predicts the central density and under-predicts the density in
the area near the radial critical curve. It even suggests a fictitious
dip in this area. When we repeat the same exercise but reducing the
number of cells down to 300 (and keeping $\epsilon = 2 \times
10^{-11}$, we observe a similar behaviour to the one described in
section \ref{sect_systemI}$^{\star}$.  The recovered 1D profile does
not show a dip at 20'' and the profile falls faster at radii larger
than 60''. Between 20'' and 60'', the 1D profile over-predicts the
real one by about 20 \% in the case with 300 cells.

An interesting lesson can be learned when we combine both results. The
recovered mass distribution interior to the radial critical curve is
closer to the real one when we use a smaller number of cells (300) but
between the radial and the tangential critical curves, the recovered
mass profile is better when we increase the number of cells
(500-600). Unsurprisingly, we are also able to conclude that we
definitely recover a biased mass distribution beyond 70''or 80'' from
the centre.

\begin{figure}
   \epsfysize=8.cm 
   \begin{minipage}{\epsfysize}
          \epsffile{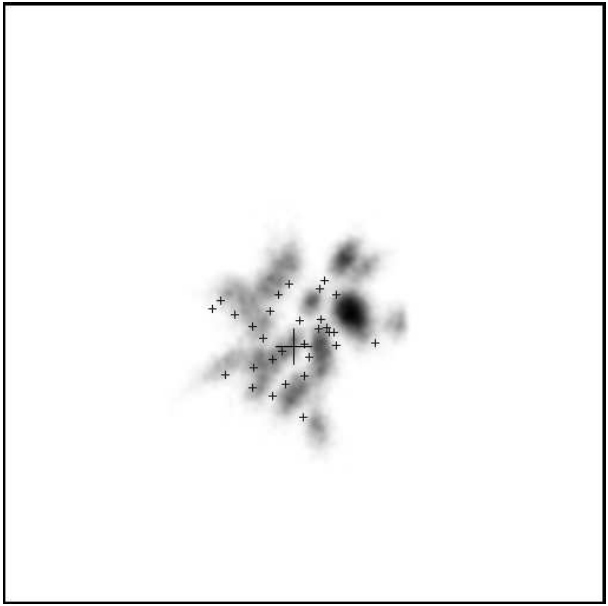}
   \end{minipage}
   \caption{
            Reconstructed positions of the sources (grey scale) for 
            1000 minimizations. The true position of the sources is marked 
            with small crosses. The big cross is the position of the main 
            halo. The field of view is a zoomed version ($2 \times 2$ arcmin) 
            of the original 5 arcmin field of view.
           }
   \label{fig_BetaReconst2}
\end{figure}

Regarding the location of the sources, the recovered $\beta$ positions
deviate from the true position by between 0'' and 5'' (see figure
\ref{fig_BetaReconst2}). Reducing the number of cells from 500 to 300
does not show any appreciable improvement in this situation and the
recovered  $\beta$  positions look almost indistinguishable from
figure \ref{fig_BetaReconst2}. This is to be contrasted by case $ii)$
in section \ref{sect_systemI}. However, as opposed to that case,
reducing the number of cells to 300 in the
simulated data does not here produce a sequence of almost identical
grids. This suggest that the recovered positions of case $ii)$ in
section \ref{sect_systemI} (see figure \ref{fig_RecovB3}) are more the
product of fixing the grid than being the real position of the
sources. % question is: why does the grid not change in that case?
         % any chance we found a ``optimal'' grid in that case? I
         % don't understand why the algorithm should ``fix'' the grid
         % for no reason?

%%%%%%%%%%%%%%%%%%%%% 
\section{Conclusions} 
%%%%%%%%%%%%%%%%%%%%%
Using a non-parametric algorithm (SLAP) we reconstruct the mass
distribution of A1689 based on strong lensing data containing the 106
multiply lensed images identified by B2004. The reconstructed mass
agrees well with previous estimations based on parametric algorithms
(B2004). Our non parametric approach is an essential complement to the
more model dependent methods and also allows us to understand better
the uncertainties and potential ambiguities involved in using strong lensing
data for generating surface mass distributions.  In particular, we
find that our recovered mass is biased toward smaller values beyond
the most external tangential arcs and there is some evidence for
degeneracy problems in the very central region. However, we also
conclude that the total mass can be well constrained within 70'' from
the center of the cluster. The total projected mass within 70'' from
the center is found to be $0.25 \times 10^{15} h^{-1} M_{\odot}$. 
The simulated work suggest that the estimated profile between 
20'' and 70'' is reliable.\\ 
It also shows how the degeneracy in the central region can be reduced by 
taking a smaller number of cells which naturally decreases 
the degrees of freedom. This is done at the expense of a bias 
in the outer regions which is increased when the number of cells 
is low. Testing the algorithm with simulations 
which mimic the real data and the average estimated mass we found 
that the best results can be obtained combining a minimization 
with a relatively large number of cells ($N_c \approx 500$) with a 
minimization with a smaller number of cells ($N_c \approx 300$). 
Combining these results we find that the mass recovered in a 
non-parametric way is compatible with a NFW profile plus an 
excess associated to substructure around the central 
overdensity. 

Our modeling indicates that the central region of the cluster is
either affected by projection along the line of sight or is not yet
fully relaxed as significant local density perturbations are found in
our reconstruction. 
%(XXX I have tended tthrought the paper to
%distinguish between dynamical evidence of a merger as may be seen in
%velocity or with X-ray info - from lensing whcih of course cannot
%directly say anything about merging - only about close projection and
%not about motion). 
Evidence of ongoing merging has been also reported from
an analysis of recent X-ray data (Andersson \& Madejski 2004).  The
mass derived from the X-ray profile is about two times smaller than
the one derived here when the cluster is assumed to be in a relaxed
state (Andersson \& Madejski 2004).  If one believes the lensing
results, it means the assumption of hydrostatic equilibrium used to
derive the mass from X-rays may be hard to justify in detail (Xue \& Wu, 2002).\\ 
Previous analysis of A1689 using different lensing techniques support this
hypothesis as they tend to agree in the mass (Tyson \& Fischer 1995,
Taylor et al. 1998, Dye et al. 2001).
Our integrated mass estimate agrees well with these previous analises$^{\star}$.  \\

In the bibliography one can find numerous studies of how 
masses derived from X-rays, optical and lensing compare 
(Miralda-Escud\'e  \& Babul 1995, Allen 1998, Wu et al. 1998, 
Wu 2000, Cypriano et al. 2004). Systematically, a discrepancy of 
about 2-4 is found in the central regions of some clusters, 
specially in the ones with evidence of being in a non-relaxed 
state (Allen 1998). A combination of the gravitational potential 
in the central region derived from strong lensing observations 
with high resolution X-ray data will allow exciting studies 
focusing on the dynamical state of the gas in these regions.
Also interesting is to combine the strong lensing results in the central 
region with weak lensing information which allows to extend the analysis  
up to Mpc scales (Broadhurst et al. 2004b).

%XXX I will add some more on this soon - we shoudl mention how we can improve on 
%this work and extend it to the weak regime to make use of the general more weakly lensing image shape information etc.

%%%%%%%%%%%%%%%%%%%%%%%%%%% 
\section{Acknowledgments} 
%%%%%%%%%%%%%%%%%%%%%%%%%%% 
This work was supported by NSF CAREER grant AST-0134999, NASA grant 
NAG5-11099, the David and Lucile Packard Foundation and  
the Cottrell Foundation. The authors would like to thank 
Elizabeth E. Brait and E.Hayashi for useful discussions.

%\newpage

%%%%%%%%%%%%%%%%%%%%%%%%%%%%%%%

%%%%%%%%%%%%%%%%%%%%%%%%%%%%%%%%%%%%%%%%%%%%%%%%%%%%%%%%%%%%%%%%%%%%%%%
%%%%%%%%%%%%%%%%%%%%%%%%%%%%%%%%%%%%%%%%%%%%%%%%%%%%%%%%%%%%%%%%%%%%%%%
%%%%%%%%%%%%%%%%%%%%%%%%%%%%%%%%%%%%%%%%%%%%%%%%%%%%%%%%%%%%%%%%%%%%%%%

\bsp
\label{lastpage}
\end{document}